\documentclass{article}

\usepackage{arxiv}

\usepackage[utf8]{inputenc} % allow utf-8 input
\usepackage[T1]{fontenc}    % use 8-bit T1 fonts
\usepackage{hyperref}       % hyperlinks
\usepackage{url}            % simple URL typesetting
\usepackage{booktabs}       % professional-quality tables
\usepackage{amsfonts}       % blackboard math symbols
\usepackage{nicefrac}       % compact symbols for 1/2, etc.
\usepackage{microtype}      % microtypography
\usepackage{lipsum}
\usepackage{fancyhdr}       % header
\usepackage{graphicx}       % graphics
\graphicspath{{Figs/}}     % organize your images and other figures under media/ folder
\usepackage{subfigure}
\usepackage{amsmath,amssymb,amsfonts}
\usepackage{multirow}
\usepackage{xcolor}

\title{Empowering Tuberculosis Screening with Explainable Self-Supervised Deep Neural Networks}
\author{
  Neel Patel \\
  Department of Mechanical and Mechatronics Engineering \\
  University of Waterloo \\
  Waterloo, ON N2L 3G1, Canada\\
  \And
  Alexander Wong \\
  Department of Systems Design Engineering \\
  University of Waterloo \\
  Waterloo, ON N2L 3G1, Canada \\
 \And
  Ashkan Ebadi \\
  Digital Technologies Research Centre \\
  National Research Council Canada \\
  Toronto, ON M5T 3J1, Canada\\
  \texttt{ashkan.ebadi@nrc-cnrc.gc.ca} \\
}

\begin{document}
\maketitle

% Abstract (Do not insert blank lines, i.e. \\) 
\begin{abstract}
Tuberculosis persists as a global health crisis, especially in resource-limited populations and remote regions, with more than 10 million individuals newly infected annually. It stands as a stark symbol of inequity in public health. Tuberculosis impacts roughly a quarter of the global populace, with the majority of cases concentrated in eight countries, accounting for two-thirds of all tuberculosis infections. Although a severe ailment, tuberculosis is both curable and manageable. However, early detection and screening of at-risk populations are imperative. Chest x-ray stands as the predominant imaging technique utilized in tuberculosis screening efforts. However, x-ray screening necessitates skilled radiologists, a resource often scarce, particularly in remote regions with limited resources. Consequently, there is a pressing need for artificial intelligence (AI)-powered systems to support clinicians and healthcare providers in swift screening. However, training a reliable AI model necessitates large-scale high-quality data, which can be difficult and costly to acquire. Inspired by these challenges, in this work, we introduce an explainable self-supervised self-train learning network tailored for tuberculosis case screening. The network achieves an outstanding overall accuracy of 98.14\% and demonstrates high recall and precision rates of 95.72\% and 99.44\%, respectively, in identifying tuberculosis cases, effectively capturing clinically significant features.
\end{abstract}

% Keywords
\keywords{Tuberculosis \and Deep learning \and Explainable neural network \and Rapid screening \and Radiology}

\section{Introduction}
% Background on Tuberculosis (TB)
Tuberculosis (TB), caused by the transmission of the bacillus Mycobacterium tuberculosis through airborne particles expelled by individuals with the illness \cite{who2023}, is estimated to have infected approximately a quarter of the world's population \cite{un2022}, especially in regions grappling with poverty and economic hardship \cite{who2023}. TB is a disease that can typically be prevented and cured. However, in 2022, it stood as the second most fatal infectious disease globally, following the coronavirus disease (COVID-19), and accounted for nearly double the number of fatalities compared to human immunodeficiency virus/acquired immunodeficiency syndrome (HIV/AIDS) \cite{who2023}. After infection, the highest risk of TB disease occurs within the initial two years ($\approx 5\%$), decreasing significantly thereafter \cite{un2022}. And, some individuals will completely clear the infection \cite{emery2021}. Among those who develop TB annually, around 90\% of cases occur in adults, with a greater prevalence observed among men \cite{who2023}. Although tuberculosis is treatable, with about 85\% of infections effectively cured using a six-month antibiotic regimen \cite{wong2022}, the mortality rate from untreated TB disease remains notably high at around 50\% \cite{tiemersma2011}.

% Emphasis on the importance of early detection and treatment for effective disease management. 
Screening high-risk populations and early disease detection are vital steps in TB treatment \cite{wong2022}, but tuberculosis continues to be either underdiagnosed or underreported to national authorities \cite{who2023}. The situation worsened during the COVID-19 pandemic as evidenced by a significant global decline in newly diagnosed and officially reported TB cases, indicated by an 18\% reduction between 2019 and 2020, after substantial increases from 2017 to 2019, dropping from 7.1 million to 5.8 million cases, with a partial recovery to 6.4 million in 2021 \cite{who2023}. In 2022, the global count of newly diagnosed TB cases, officially reported and notified, reached 7.5 million, marking the highest count since the inception of global TB monitoring by the World Health Organization (WHO) in 1995 \cite{who2023}. Moreover, the global decrease in TB-related deaths from 2015 to 2022 amounted to 19\%, falling considerably short of the WHO's End TB Strategy target of a 75\% reduction by 2025 \cite{who2023}.

Whether it is improved medication and care or more effective resistance management, the cornerstone of success in TB treatment lies in early and accurate diagnostics \cite{wong2022,nema2012}. As the predominant modality utilized in tuberculosis screening, chest x-ray (CXR) imaging has proven to be highly efficient and cost-effective \cite{diaz2020,li2018}. However, it poses further challenges due to the presence of atypical radiographic presentation and shortages of radiologists \cite{chexaid2020}, especially in resource-limited settings, since it necessitates skilled human readers or trained clinicians/technicians for interpretation \cite{wong2022}.

Due to the global scarcity of experienced individuals for interpreting CXR images in tuberculosis screening, there has been a notable surge in interest in artificial intelligence (AI)-driven TB screening solutions, e.g., \cite{rahman2020,singh2019,wong2022}. AI-driven systems hold promise in healthcare by offering efficient analysis of vast amounts of medical data, aiding in diagnosis, treatment planning, and personalized care, ultimately improving patient outcomes and streamlining healthcare delivery \cite{sunarti2021}. AI-powered TB screening can assist clinicians by providing efficient and reliable analysis of imaging data, thereby optimizing resource utilization and enabling early detection of the disease. Indeed, the latest report from the WHO highlights further products being considered for review, including point-of-care TB tests and computer-aided detection (CAD) for digital chest radiography in individuals under 15 years of age, among other potential applications \cite{who2023}. 

Inspired by the urgent demand in resource-limited settings and low-income populations and the WHO's recent endorsement of CAD for tuberculosis screening, we present an explainable self-supervised deep neural network tailored for tuberculosis case screening. We leveraged a framework named distillation for self-supervision and self-train learning (DISTL) \cite{park2022} that is inspired by the learning process of radiologists. DISTL incorporates both self-supervision and self-training through knowledge distillation, enabling the model to learn from unlabeled data and iteratively improve its performance. The approach overcomes the need for large-scale high-quality data in building reliable medical AI models by utilizing limited labelled and extensive unlabeled data, mirroring the teacher-student learning paradigm. We also conducted explainability analysis to ensure the network effectively identifies disease-related patterns and indicators in the CXR images. This research holds significant promise in addressing current challenges and we hope it could revolutionize TB diagnosis through its innovative blend of explainable self-supervised learning and deep neural networks. 

The structure of the article is outlined as follows: Section~\ref{sec:methods} elaborates on the data and methodology. Section~\ref{sec:results} showcases the study's findings, while Section~\ref{sec:discuss} offers conclusions and discussions on broader implications. Lastly, Section~\ref{sec:limit} outlines the study's limitations and suggests future directions.

%%%%%%%%%%%%%%%%%%%%%%%%%%%%%%%%%%%%%%%%%%
\section{Data and Methods}\label{sec:methods}
This section provided a comprehensive overview of the data and methods utilized in the study, detailing data preparation, network architecture, and the validation process driven by explainability analysis.

\subsection{Data Collection and Preparation}
We used a comprehensive chest x-ray (CXR) dataset comprising four distinct sources to ensure diversity and robustness. Tuberculosis-positive images ($n=2,141$) were sourced from the Montgomery ($n=58$) \cite{montgomery2014}, Shenzhen ($n=336$) \cite{montgomery2014}, Belarus ($n=1,047$) \cite{belarus}, and Rahman et al. ($n=700$) \cite{rahman2020} datasets. Additionally, the normal images ($n=3,500$) were exclusively collected from Rahman et al. \cite{rahman2020} dataset. 

After data collection and integration, we proceeded with several preprocessing steps. Initially, we employed the U-Net model \cite{unet2015} to extract the lung region from the CXR images. The U-Net architecture is specifically designed to handle biomedical images and has gained widespread recognition in medical image segmentation \cite{asgari2021}. For instance, in \cite{wang2017}, a U-Net-based image segmentation approach demonstrated superior performance compared to traditional methods in segmenting lung, heart, and clavicle structures in chest radiographs. U-Net effectively manages the spatial complexities of medical images by employing a contracting path to capture contextual information and a symmetric expanding path to enable precise localization. The U-Net model predicted the lung regions, producing segmentation masks. These masks were then resized to match the original dimensions of CXRs, using nearest-neighbor interpolation, to preserve the aspect ratio of the lung regions. The x-ray images were cropped using the two largest contours identified. The segmentation step ensured that subsequent analyses focused solely on the relevant lung regions, effectively excluding any extraneous artifacts, text, or markers present in the original images. The cropped images were then resized to 225x225 pixels. 

\begin{figure}
    \centering
    \subfigure[Normal case, original]
    {
        \includegraphics[scale=0.18]{"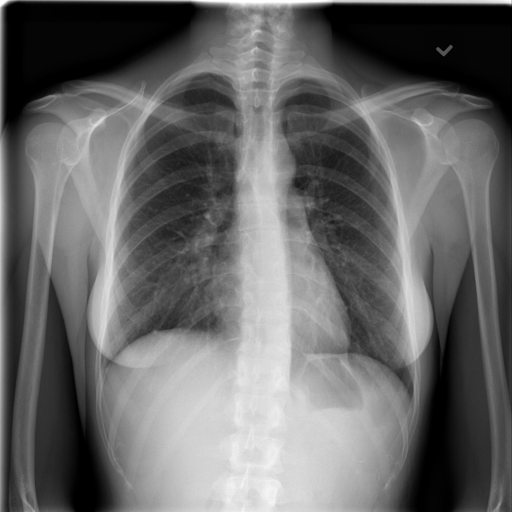"}
    }
    \subfigure[Normal case, cropped]
    {
        \includegraphics[scale=0.405]{"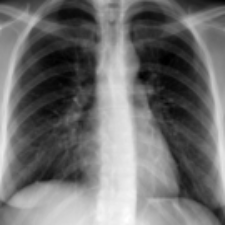"}
    }
    \subfigure[TB case, original]
    {
        \includegraphics[scale=0.0185]{"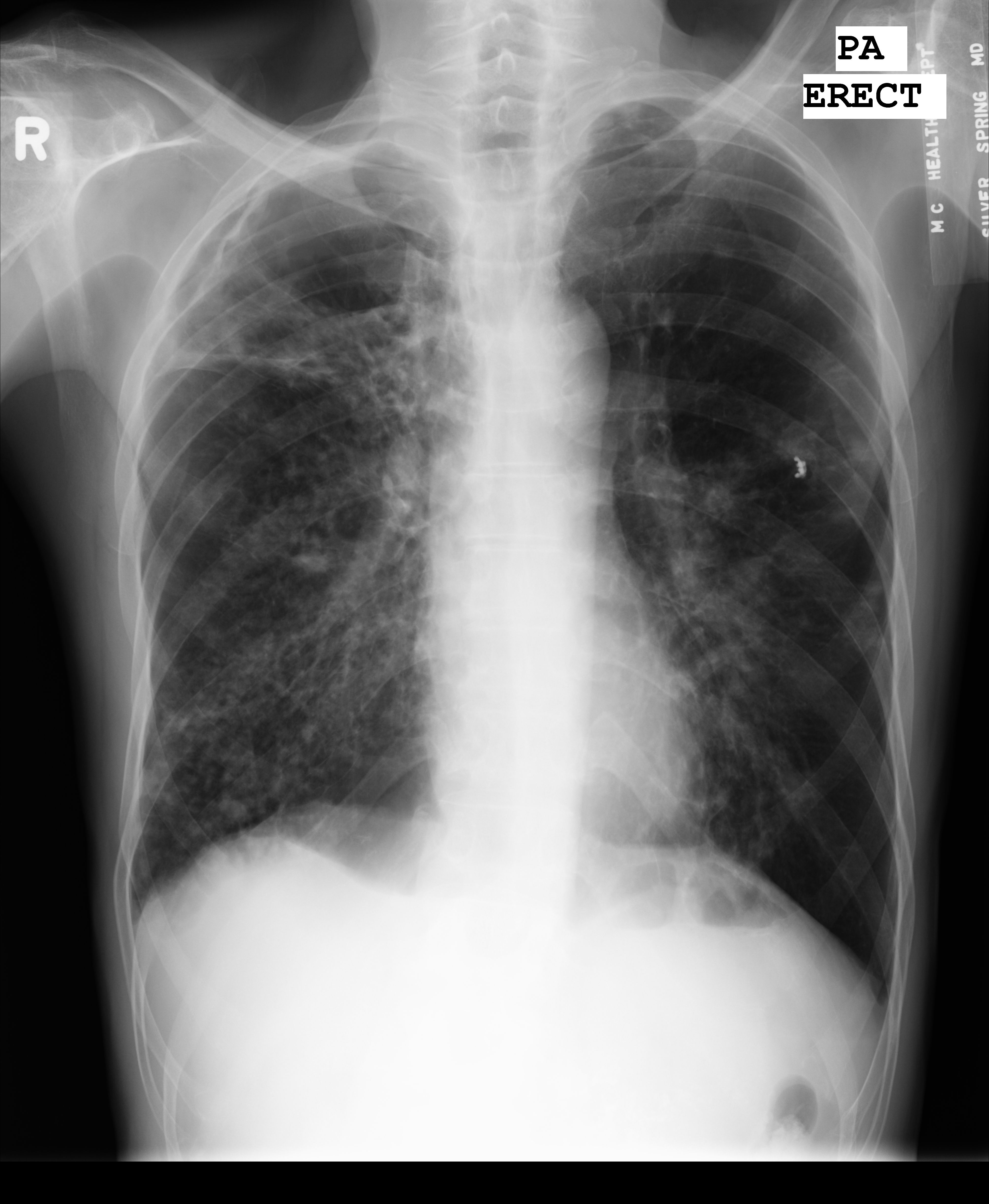"}
    }
    \subfigure[TB case, cropped]
    {
        \includegraphics[scale=0.4]{"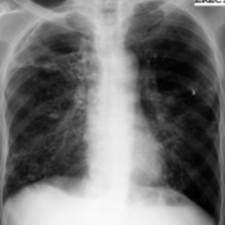"}
    }
    \caption{Sample images in the dataset.}
    \label{fig1}
\end{figure}

Next, we conducted a manual review to identify and exclude images exhibiting poor segmentation outcomes, such as those depicting only one lung or misidentified non-lung regions. This review led to the removal of 516 normal and 218 tuberculosis images that did not meet quality standards. The final dataset consisted of $1,923$ tuberculosis-positive and $2,984$ normal images. The dataset was subsequently partitioned into training and test subsets, with 10\% of the images reserved for testing and evaluating the model's performance, leaving the remaining 90\% for training purposes. Upon partitioning the dataset, the training subset was further divided into four portions. Specifically, 10\% of the training data was designated as labeled data, while the remaining 90\% was designated as unlabeled data and evenly distributed among three subsets. Figure~\ref{fig1} depicts sample normal and tuberculosis images, original and segmented, in our dataset.

\subsection{Network Architecture}
Our tuberculosis screening network adopts the distillation for self-supervision and self-train learning (DISTL) framework \cite{park2022}. DISTL draws inspiration from the learning process of radiologists, enhancing the performance of vision transformers by incorporating self-supervision and self-training techniques through knowledge distillation simultaneously \cite{park2022}. In the self-training approach \cite{xie2020}, a learner (referred to as the \textit{teacher}), initially trained with limited labeled data, continues to label a large pool of unlabeled data, creating pseudo-labels. These pseudo-labels are then utilized to train a new model (referred to as the \textit{student}) with an expanded dataset. This teacher-student learning framework is commonly known as knowledge distillation \cite{park2022}. 

In our self-supervised self-train learning framework, both the teacher and student models utilize the ViT small model, as the backbone of the network, with the student model benefiting from a drop path for improved regularization. The ViT small model \cite{park2022} that we employed was pre-trained on the CheXpert dataset \cite{chexpert2019}, tailored for CXR analysis. Specifically, the model was pre-trained on five radiological categories: lung opacity, consolidation, edema, pneumonia, and pleural effusion. These categories were used for identifying various manifestations of infectious diseases. Through training on these specific categories and radiological markers, the model has been finely tuned to improve its performance and adaptability, effectively managing a diverse range of patient conditions and varying imaging settings.

Images are partitioned into $8x8$ pixel patches, converting each patch into a $384$-dimensional embedding vector to capture local features within chest x-rays. This high-dimensional embedding is essential for capturing complex patterns indicative of pathological changes. Additionally, the network consists of 12 transformer layers with 6 attention heads each, augmenting its ability to detect dependencies among different regions of the image. Layer normalization ($\epsilon=1\text{e-}6$) was employed to ensure stability. Subsequently, the models were encapsulated within a multi crop wrapper (MCW) to accommodate inputs of varying resolutions.

\begin{figure}
\centerline{\includegraphics[width=0.8\textwidth]{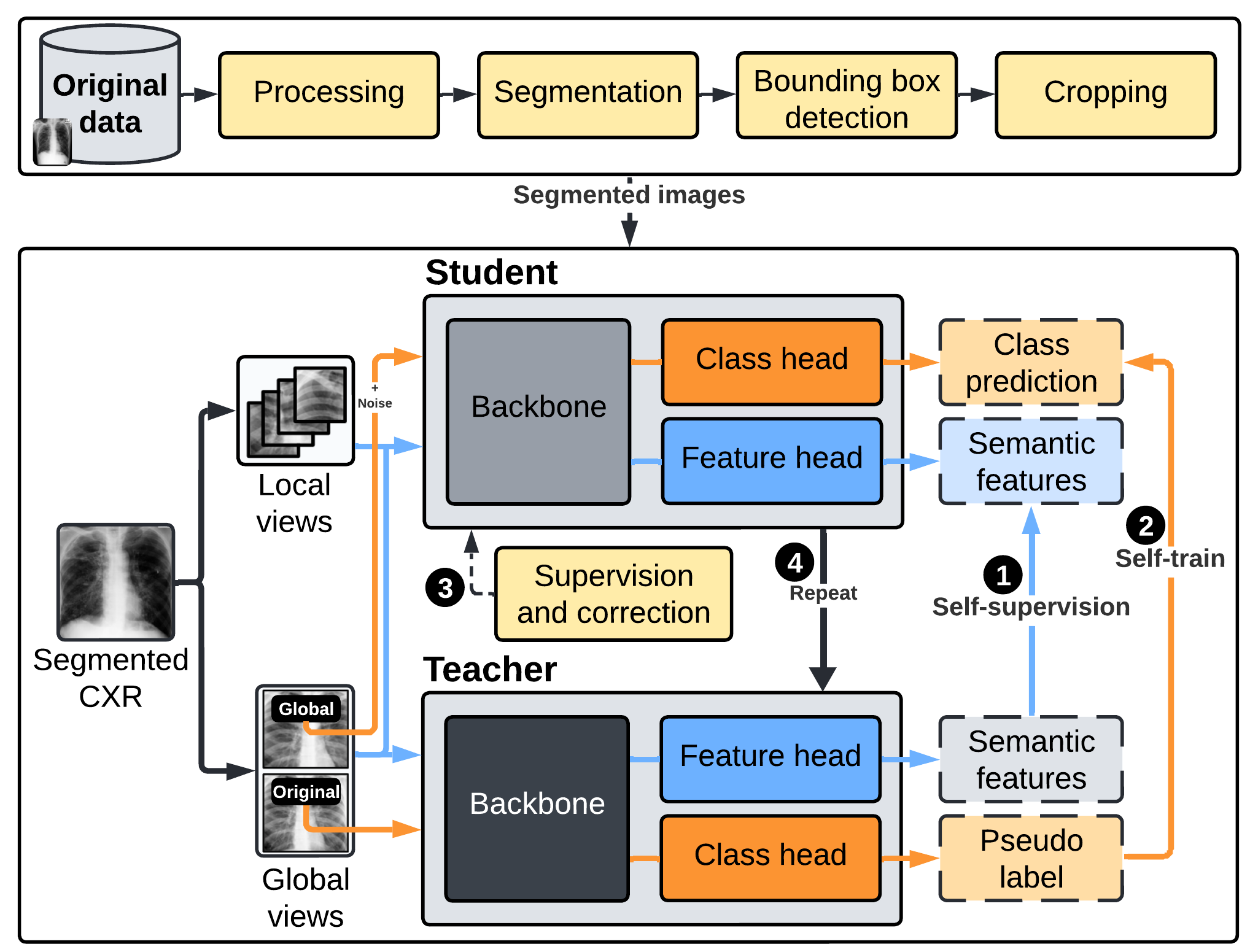}}
\caption{The high-level conceptual flow of the framework that encompasses three primary components for segmentation, self-supervision, and self-training.}
\label{fig2}
\end{figure}

The network incorporates two distinct multilayer perceptrons (MLP) heads: 1) The self-distillation with no labels (DINO) head \cite{dino2021}: Integrated as a feature generation head within the DISTL framework for self-supervised learning, the DINO head utilizes batch normalization and the Gaussian error linear unit (GELU) activation to produce normalized feature vectors. These features support contrastive loss calculations, promoting robust, label-independent learning by enhancing feature discriminability and stability across different perspectives. 2) The classifier (CLS) head: It is attached directly to the ViT model and is tasked with binary classification of the processed images, determining the presence or absence of tuberculosis. The classification module comprises sequential linear layers followed by rectified linear unit (ReLU) activations, mapping the dense transformer outputs to the final classification task. The high-level conceptual flow of the framework is depicted in Figure~\ref{fig2}.

% figure of the network architecture and flow

\subsection{Network Training and Performance Evaluation}
Feature extraction was conducted using the DINO head, and tuberculosis detection was performed using the CLS head. We adapted the weights from the CheXpert pre-trained model and made minor adjustments to ensure compatibility with our training environment. We applied various data augmentation techniques to the original dataset, including random resizing, cropping, color jittering, rotation, auto contrast, equalization, and blurring, to prepare it for robust training across diverse conditions. The student model learned from labeled data by minimizing the binary cross entropy (BCE) with logits loss (BCEWithLogitsLoss) between its predictions and the labels derived from the transformed images. The loss function measured the error between the student's predictions and the actual labels, directing backpropagation and parameter adjustments through the AdamW optimizer, which dynamically adjusts the weights based on cosine annealing schedules for both the learning rate and weight decay. Mixed precision training was utilized to boost both performance and efficiency.

For unlabeled training, both teacher and student models were initialized with matching weights extracted from the saved state dictionary of the student model, ensuring structural uniformity. This method guarantees that both models begin with a resilient, pre-trained base encompassing learned patterns specific to tuberculosis. A three-tiered augmentation strategy was employed comprising two global and one local augmentation methods. This configuration aimed to offer diverse perspectives of the input images, fostering robust feature learning during the self-supervised phase of training. The global part ranged from minimal to extensive augmentation (e.g., rotation, auto contrast) to emulate various viewing conditions and imaging variances observed in clinical scenarios. Conversely, the local part included intensive augmentation to capture intricate details, which are crucial for discerning subtle pathological characteristics in CXR images.

There were a total of three runs utilizing the unlabeled training data portions, with each new run incorporating an additional subset of unlabeled data. As training progressed through these subsequent runs, the student model was continually initialized from its updated state dictionary. In contrast, the teacher model was loaded from its own state dictionary, potentially containing more generalized and stable knowledge accumulated over multiple iterations. This enabled the teacher model to serve as a consistent and comprehensive guide, assisting the student model in stabilizing its learning amidst the increasing complexity of the data (Figure~\ref{fig2}). In other words, with each subsequent training run, a larger portion of the unlabeled dataset was utilized. This gradual expansion of data exposure ensured that the models were not overwhelmed prematurely and allowed them to learn from intricate, unlabeled inputs, thereby enhancing their performance.

The training process incorporated two primary loss functions: 1) DINOLoss: This loss function plays a crucial role in the self-supervised learning aspect of the DISTL framework. In this setup, the teacher model processes only the global views of the images, which are expected to capture more general features, while the student model receives both global and local views (see Figure~\ref{fig2}). This design encourages the student to learn from a richer, more varied context. DINOLoss is designed to minimize the difference in responses between the teacher and the student on the global views, thus encouraging the student model to internalize and replicate the semantic features perceived in the global crops. This mechanism aims to enhance the student’s ability to generalize from broad visual cues without relying directly on labeled data. 2) BCEWithLogitsLoss: It was employed for the self-training part of the training process. This loss function was used to align the student's classifications (computed from both global and local views) with the teacher's predictions (derived from the global views). This loss was calculated by comparing the transformed outputs from the teacher with the corresponding outputs from the student, adjusting the student’s understanding to be more in line with the teacher’s perspective. This alignment helps refine the student’s predictive capabilities on the task at hand. Both losses were weighted and combined to achieve a balance between self-supervised and supervised learning, enabling flexible adjustment of learning priorities throughout training epochs.

The following strategy was proceeded for parameter updating. For the student model, backpropagation was used to update it with gradients computed from the combined loss. For the teacher model, updates were performed through an exponential moving average (EMA) of the student model’s parameters, integrating refined student parameters over time to stabilize learning. Periodic adjustments were also implemented. Every 500 iterations, comparisons against the student model were made using labeled data. This was accomplished through BCEWithLogitsLoss computation, comparing the student's classification outputs with the labels. This supervised adjustment fine-tuned the student model's performance, which was crucial for maintaining accuracy.

The model performance was evaluated using several metrics including recall, precision, and accuracy scores. These metrics assessed the diagnostic performance of the model in distinguishing TB-positive and -negative cases. Furthermore, we compared the performance of our model against four fine-tuned convolutional neural network (CNN)-based baseline models, namely a custom vanilla CNN, VGG16, ResNet18, and ResNet50. All experiments and evaluations were carried out within an environment with PyTorch version 2.2.1+cu118, Pillow version 9.0.1, OpenCV-Python version 4.9.0.80, and Scikit-Learn version 1.0.2. These components were operated on a system equipped with CUDA 11.1 and an NVIDIA RTX 4090 GPU, running Python version 3.8.0.

\subsection{Explainability Analysis}
DISTL provides a simpler localization of lesions through the model's attention mechanism. We adopted the methodology outlined in \cite{park2022} to evaluate whether the network effectively identifies disease-related patterns or indicators in the CXR images. It is argued that ViT achieves superior localization compared to the CNN due to its direct attention mechanism, as opposed to the CNN's indirect attention, e.g., through Gradient-weighted Class Activation Mapping (GradCAM) \cite{park2022}. Hence, we assessed the localization performance with model attention. The predictions generated by model attention were derived by applying threshold values post-normalization to localize the target lesions. Given the ViT model's multiple heads available for visualization, the best-performing head was chosen for evaluation and visualization purposes.

%%%%%%%%%%%%%%%%%%%%%%%%%%%%%%%%%%%%%%%%%%
\section{Results}\label{sec:results}
We assessed the effectiveness of our network in detecting TB cases from CXR images through two approaches. First, we evaluated the network's quantitative performance and compared it against various baseline models. Subsequently, we examined its decision-making process by conducting an explainability performance validation. Further analysis of the results will be elaborated upon in this section.

\subsection{Performance Analysis}
Table~\ref{tab1} presents the performance comparison results between our self-supervised self-trained network and four baseline models. All five models underwent training and testing on identical datasets. As expected, the vanilla CNN model, characterized by a straightforward architecture comprising two convolutional layers and a fully connected layer, exhibits the lowest recall for TB detection. The VGG and ResNet architectures demonstrate enhancements over the vanilla CNN. The VGG16 model achieves an accuracy of 96.07\%, whereas ResNet18 and ResNet50 attained scores of 96.49\% and 97.31\%, respectively. A consistent trend of escalating precision and recall in both categories is apparent across these models, with the ResNet50 model surpassing the others in the TB class. As observed, our model, employing image segmentation and self-attention techniques, outperforms all other models across all classes and performance metrics. It achieves an overall accuracy of 98.14\% and demonstrates exceptional precision and recall in both normal and tuberculosis classes. The process of distilling knowledge through self-supervised learning and self-training, despite the lack of lesion-specific information, is argued to foster a robust correlation between attention and the lesion, which can potentially enhance the model's diagnostic accuracy \cite{park2022}.

\begin{table}
 \caption{Model performance comparison.}
  \centering
  \begin{tabular}{llllll}
    \toprule
    \textbf{Model}	& \textbf{Class}	& \textbf{Precision}	& \textbf{Recall}	& \textbf{F1 score}	& \textbf{Accuracy}\\
    \midrule
    \multirow{2}{*}{\textbf{Vanilla CNN}} & \textbf{N} & 92.11\% & 98.32\% & 95.11\% & \multirow{2}{*}{93.80\%}  \\
    \cline{2-5}
    \textbf{} & \textbf{TB} & 97.01\% & 86.63\% & 91.53\% &  \\
    \hline
    \multirow{2}{*}{\textbf{VGG16}} & \textbf{N} & 94.84\% & 98.99\% & 96.87\% & \multirow{2}{*}{96.07\%} \\
    \cline{2-5}
    \textbf{} & \textbf{TB} & 98.28\% & 91.44\% & 94.74\% &  \\
    \hline
    \multirow{2}{*}{\textbf{ResNet18}} & \textbf{N} & 94.87\% & \textcolor[HTML]{45a831}{\textbf{99.66\%}} & 97.21\% & \multirow{2}{*}{96.49\%}  \\
    \cline{2-5}
    \textbf{} & \textbf{TB} & 99.42\% & 91.44\% & 95.26\% &  \\
    \hline
    \multirow{2}{*}{\textbf{ResNet50}} & \textbf{N} & 96.71\% & 98.99\% & 97.84\% & \multirow{2}{*}{97.31\%}  \\
    \cline{2-5}
    \textbf{} & \textbf{TB} & 98.33\% & 94.65\% & 96.46\% &  \\
    \hline
    \multirow{2}{*}{\textbf{Our model}} & \textbf{N} & \textcolor[HTML]{45a831}{\textbf{97.37\%}} & \textcolor[HTML]{45a831}{\textbf{99.66\%}} & \textcolor[HTML]{45a831}{\textbf{98.50\%}} & \multirow{2}{*}{\textbf{98.14\%}}  \\
    \cline{2-5}
    \textbf{} & \textbf{TB} & \textcolor[HTML]{9c1426}{\textbf{99.44\%}} & \textcolor[HTML]{9c1426}{\textbf{95.72\%}} & \textcolor[HTML]{9c1426}{\textbf{97.55\%}} &  \\
    \bottomrule
    \multicolumn{6}{l}{$^{\mathrm{a}}$\textbf{N:} Normal, \textbf{TB:} Tuberculosis.} \\
    \multicolumn{6}{l}{$^{\mathrm{b}}$Highest values of performance metrics for normal and tuberculosis } \\ \multicolumn{6}{l}{$^{}$ classes are highlighted in bold \textcolor[HTML]{45a831}{\textbf{green}} and \textcolor[HTML]{9c1426}{\textbf{red}}, respectively.}
  \end{tabular}
  \label{tab1}
\end{table}

\subsection{Explainability Analysis}
Radiologists rely on several critical indicators to detect tuberculosis in CXR images. Specific abnormalities observed on chest x-rays, such as upper lobe infiltrates or consolidation, cavity formation, rounded densities in lung parenchyma, pleural effusion, and bilateral hilar lymphadenopathy, strongly indicate active tuberculosis \cite{kamran2020}. These indicators, along with clinical history and other diagnostic tests, aid radiologists in the accurate detection and diagnosis of tuberculosis from CXR images.

\begin{figure}
    \centering
    \subfigure[TB case 1, segmented image]
    {
        \includegraphics[width=1.6in]{"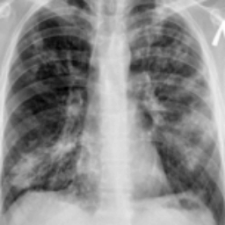"}
        \label{fig3a}
    }
    \subfigure[TB case 1, heatmap]
    {
        \includegraphics[width=1.6in]{"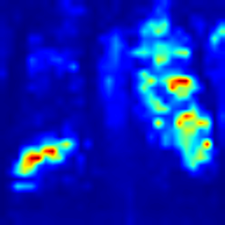"}
        \label{fig3b}
    }
    \\
    \subfigure[TB case 2, segmented image]
    {
        \includegraphics[width=1.6in]{"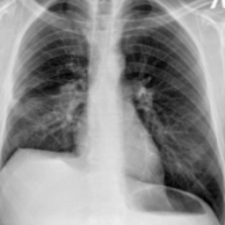"}
        \label{fig3c}
    }
    \subfigure[TB case 2, heatmap]
    {
        \includegraphics[width=1.6in]{"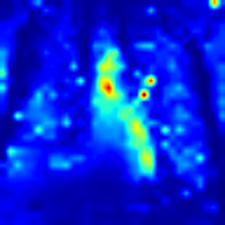"}
        \label{fig3d}
    }
    \caption{Two sample tuberculosis-positive cases, correctly classified by the proposed network, with identified critical regions.}
    \label{fig3}
\end{figure}

In addition to the quantitative performance evaluation, a thorough explainability analysis was carried out on the proposed network. Figure~\ref{fig3} illustrates two samples of TB-positive patient cases, highlighting the critical factors identified. It is evident that the model primarily relies on clinically relevant areas of the lung in the CXR images to guide its decision-making process.

\textbf{TB case 1:} As depicted in Figure~\ref{fig3b}, the heatmap highlights the superior lobe in the left lung and the middle and inferior lobes in the right lung. Abnormal densities in the lung fields, such as infiltrates or consolidations, are also visible and identified. These are consistent with observations in clinical studies \cite{koh2010} and underscore the network's ability to recognize patterns indicative of tuberculosis.

\textbf{TB case 2:} As seen in Figure~\ref{fig3d}, the heatmap demonstrates a pronounced emphasis on the superior lobe in the left lung, indicating a possible consolidation or cavitation, which are radiographic indicators of tuberculosis \cite{koh2010}. This attention pattern reaffirms the model's concordance with established radiological knowledge, where the presence of cavitary lesions and consolidation typically indicates tuberculosis.

%%%%%%%%%%%%%%%%%%%%%%%%%%%%%%%%%%%%%%%%%%
\section{Discussion and Conclusion}\label{sec:discuss}
Many AI models have been proposed in the literature for medical imaging with most of them being heavily dependent on the availability of massive labeled data of high quality \cite{park2022}. Although vast amounts of medical imaging data are accumulated annually, leveraging this data with common supervised learning approaches is hindered by label scarcity \cite{park2022}. In this work, we utilized a self-supervised, self-trained deep neural network architecture tailored for tuberculosis case screening and detection from CXR images. Our evaluation encompassed a thorough performance assessment, incorporating an explainability validation to scrutinize and authenticate the decision-making processes of the network. In experimental results, it is evident that our model excels in detecting tuberculosis cases with high performance while also displaying clinically relevant behavior.

As indicated in Table~\ref{tab1}, our model attained an impressive overall accuracy of 98.14\%. Moreover, it exhibited high recall and precision rates of 95.72\% and 99.44\% in detecting TB cases and 99.66\% and 97.37\% in identifying normal cases. When compared to the baseline models, the proposed model also demonstrated superior performance across all the performance metrics and all classes. This difference in performance can be attributed to the architectural advantages of the self-supervised self-trained model. Unlike other models reliant on convolutional layers, ours interpreted inputs as arrays of patches, employing self-attention mechanisms to capture interdependencies across the image. This approach is especially advantageous in medical imaging, where a holistic grasp of entire images is critical for accurate diagnostics. Furthermore, we leveraged multiple techniques such as multi-crop strategies to enhance the model's exposure to diverse image presentations during training. Additionally, we conducted a thorough explainability analysis to verify the network's ability to capture clinically relevant indicators of tuberculosis. As illustrated in the examples presented in Figure~\ref{fig3}, the network primarily relies on clinically relevant lung areas in CXR images to inform its decision-making process and avoids relying on erroneous visual indicators and imaging artifacts. 

In conclusion, our study presents a promising framework for tuberculosis screening, demonstrating robust performance and reliable decision-making behavior. Our team is continuously working to further validate these findings across diverse patient populations and clinical settings, paving the way for broader adoption and impact in the fight against tuberculosis. We aspire for this work to contribute to the advancement of this field, aiding researchers and clinicians in addressing the global public health crisis effectively. Additionally, we hope for this research to enhance healthcare quality for individuals experiencing poverty and economic hardship, thereby addressing significant resource limitations they encounter.

\section{Limitations and Future Work}\label{sec:limit}
This study represents an ongoing research effort, with avenues for further exploration and refinement in addressing the limitations outlined in this section. Our main objective in this work was to advance research in combating tuberculosis as a worldwide public health emergency and it is crucial to emphasize that the proposed network is not yet a fully deployable and production-ready solution. Access to patient demographics and detailed chest X-ray characteristics was unavailable to us during the study as data were acquired from publicly available data sources. We employed 1,923 TB-positive and 2,984 normal CXRs, dividing them into small labeled and large unlabeled subsets. Larger datasets are could be employed for conclusive results, suggesting a direction for future research. We conducted an explainability analysis to verify the model's identification of clinically relevant features. Moving forward, we intend to enhance the evaluation by seeking validation from certified radiologists.

\bibliographystyle{unsrt}  
\bibliography{main}  

\end{document}